\documentclass[12pt,onecolumn]{revtex4}
\usepackage{amsmath}
\usepackage{amssymb}
\usepackage{graphicx}

\newcommand{\br}{\mathbf{r}}

\begin{document}

\title{Generalisation of the fractal Einstein law relating conduction and diffusion on networks}

\author{Christophe P.\ Haynes and Anthony P.\ Roberts}
\affiliation{
School of Mathematics and Physics, The University of Queensland, Brisbane 4072, Australia
}


\begin{abstract}
In the 1980s an important 
goal of the emergent field of fractals was to determine the relationships between their physical and geometrical properties.  The fractal-Einstein and Alexander-Orbach laws, which interrelate electrical, diffusive and fractal properties, are two key theories of this type. Here we settle a long standing controversy about their exactness by showing that the properties of a class of fractal trees violate both laws. A new formula is derived which unifies the two classical results by proving that if one holds, then so must the other, and resolves a puzzling discrepancy in the properties of Eden trees and diffusion limited aggregates. The failure of the classical laws is attributed to anisotropic exploration of the network by a random walker. The occurrence of this newly revealed behaviour means that numerous theories, such as recent first passage time results, are restricted to a narrower range of networks than previously thought.
\end{abstract}

\maketitle

Consider two simple experiments performed on an arbitrary network of sites linked by bonds of identical length. First, release a vast number of random walkers at the centre and measure their average distance from the origin $\langle r \rangle$. Then replace the bonds by resistors, apply a unit voltage at a point, and earth all of the sites on a sphere centred at that point. The current gives the electrical resistance $\rho$ of the network. Although simple to conceive, the resistance $\rho$ and distance $\langle r \rangle$  are fundamental properties: in addition to quantifying mass and electronic transport in materials\cite{Havlin,Sahim,Condamin2,Gallos,Song,OrbachRev,Toulouse,Hughes2,Avra,MeakinDLA,Orbach}, they can be linked to a variety of problems such as oil recovery in porous rocks\cite{Sahim}, chemical reaction rates\cite{Condamin2} and cellular processes \cite{Gallos,Song}. These elementary properties are also connected to first passage times on networks, which have been used to model processes as diverse as viral infections and animal foraging strategies~\cite{Condamin2}.

A ubiquitous feature of fractals is that their properties follow power laws with non-integer exponents~\cite{Havlin,Mandelbrot}. For example, the mass within a radius $r$ scales as $M(r)\sim r^{d_f}$ ;  the distance travelled by a random walker scales with time as $\langle r \rangle \sim t^{1/d_w}$; the probability that a random walker is at its origin scales as $c(0,t)\sim t^{-\bar{d}/2}$ ; and the electrical resistivity between two points scales as $\rho(r) \sim r^{\zeta}$.  The exponent $d_f$ is  the fractal dimension,  $d_w$ is known as the random walk dimension, $\bar{d}$ the spectral dimension, and $\zeta$ the resistivity exponent.

The interrelationships between these exponents are the 'structure-property correlations' for fractal networks. The Alexander and Orbach\cite{Orbach} law states that $\bar{d} = 2d_f/d_w$  and Rammal and Toulouse~\cite{Toulouse} predicted that  $\zeta = d_f(2-\bar{d})/\bar{d}$.  Combining both results gives the fractal Einstein law\cite{Havlin} $\zeta = d_w-d_f$, so-called because it can also be derived from a result due to Einstein\cite{Havlin}.
Although there is preponderance of evidence in their favour, the exactness of both formulae is controversial\cite{Hughes2,Avra}: computations on two important fractals appear to violate the laws. 
Jacobs et al\cite{Jacobs} have used simulation to show that three-dimensional DLA has $\bar{d} \approx 1.35$ , whereas $2d_f/d_w \approx 1.55$.  Furthermore, their results predict $\zeta = d_w - d_f \approx 0.71$,  which disagrees with the estimated value \cite{Avra,Witten} $\zeta \approx 1$.  
The explanation of these exceptions is a long-standing challenge in the field \cite{Avra}.

\begin{figure}
\includegraphics[width = 0.6\textwidth]{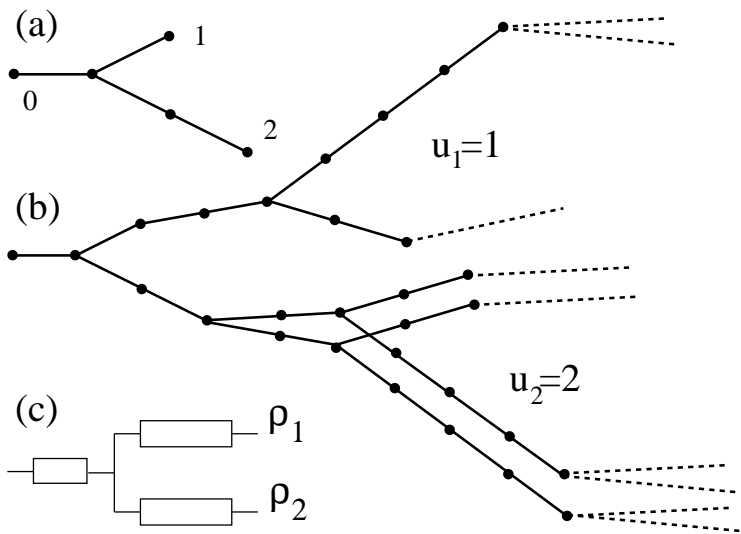}
\caption{A model fractal tree. (a) The asymmetric base unit; (b) the second generation of the network obtained by attaching multiple copies of the rescaled base unit to the end points (1) and (2) of the base unit; (c) an equivalent resistor network used to calculate $\rho^*$.}
\label{tree}
\end{figure} 
To investigate this problem we have studied the properties of a class of fractal trees \cite{Doyle,Burioni3,Kron,Tony}, an example of which is shown in Figure 1.  The network is made by taking a base unit, doubling its size, and attaching $u_i$ ($i=1,2$) copies of the re-scaled unit to each of the two end points of the base.  Continuing the process indefinitely gives an infinite network with $d_f = \ln(2(u_1+u_2))/\ln(2)$ and $d_w= 2$.  To find the resistance $\rho^*$   between the origin and "infinity"\cite{Doyle}, we represent the infinite network by three resistors: the stem of length 1 and two branches of resistivity $\rho_1$   and $\rho_2$.  Kirchoff's laws states that $\rho_1 = 1 + 2\rho^*/u_1$ and $\rho_2 = 2 + 2\rho^*/u_2$  , where $2\rho^*$ is the resistance of the infinite branches connected to the two end points of the base unit.  
Because the shape of each branch is identical with the original infinite network (but each element is twice as long), their resistance is $2\rho^*$.  Kirchoff's law for the three-resistor circuit gives $\rho^* = 1 + 1/(1/\rho_1 + 1/\rho_2)$  , which is quadratic in $\rho^*$.  Renormalisation methods can be used to derive (Haynes \& Roberts, in preparation) the spectral dimension and resistivity exponent.  For $(u_1,u_2) \neq (1,1)$, the quadratic has a finite positive root, and it can be shown that
\begin{equation}
\bar{d} = \ln \left( \frac{  2 \left( \left( 2 \rho^* + 3u_2 \right)u_1 + 2\rho^* u_2\right)^2 }{u_1^2u_2 + 4\left( \rho^{*2} + 3\rho^*u_2 + u_2^2\right)u_1 + 4\rho^{*2}u_2^2}\right)/\ln(2).
\end{equation}
For the case $(u_1,u_2) = (1,1)$, the quadratic has no positive solutions (implying $\rho^*$ is infinite), and we can show $\bar{d} = \ln\left( 2\left(u_1 + u_2\right)\right)/\ln(2)$.  For all cases, the resistivity exponent is $\zeta = 2-\bar{d}$.

The properties of the network violate the Alexander-Orbach and fractal Einstein relationships unless $u_1 = u_2 = 1$  or $u_1 = 2u_2$  .  For example, $u_1 = 1$   and $u_2 = 20$ gives $\bar{d} \approx 4.37$, which does not equal $2d_f/d_w = 5.39$.  Similarly, $\zeta = 2 - \bar{d} = -2.37$  disagrees with the prediction $\zeta = d_w - d_f = -3.34$ .  Standard computations (Appx.~\ref{AppxA}) were used to verify $\bar{d}$, $d_w$, $d_f$, and $\zeta$. 

In order to derive a new relationship between the electrostatic and diffusive properties of a network, consider the concentration field generated by the release of a random walker at the origin at every time step.  This concentration is exactly given by $C(r,t) = \int_0^\infty c(r,\tau) d\tau$, where $c(r,t)$ is the probability of finding a random walker at $r$, after time $t$, if a single walker is released at the origin at $t=0$.  To link the dynamic and static problems the integration is terminated at $T = (R/b)^{d_w}$ , where $b$ is a number of order one.  As $R = bT^{1/d_w}$ is a typical distance reached by the initial walker after time $T$, only a very small proportion of the $T+1$ walkers released will exceed this radius; hence $C(r,T)\approx 0$ for $r \geq R$.  In the central region the spatial concentration profile $C(r,T)$ is assumed to have equilibrated, and therefore satisfies the potential equation.  The boundary conditions correspond to the potential on a finite network grounded at radius $R$ due to the supply of unit current at the origin $(I=1)$.  Now the resistance is simply $\rho(R) \approx C(0,(R/b)^{d_w})/I$, which implies 
\begin{displaymath}
\rho(R) \approx \int_0^{(R/b)^{d_w}} c(0,t) dt \sim \left\{ 
\begin{array}{ccc}
R^{d_w(2-\bar{d})/2} & \bar{d} < 2 \\
\log(R) & \bar{d} = 2 \\
\rho^* - QR^{d_w(2-\bar{d})/2} & \bar{d} > 2 
\end{array} \right.
\end{displaymath} 
where $Q$  is a constant.  This exactly matches the known scaling behaviour of the resistance if
\begin{equation}
\zeta = d_w(2-\bar{d})/2.
\end{equation}                    			   
Note that the spectral dimension\cite{Hattori}, and hence $\zeta$  , are site independent, even though $\rho^*$  can vary from site to site if $\bar{d}>2$ .  Equation (2) is exact for the fractal trees depicted in Fig.~1 as $d_w = 2$. Computational checks on the result are provided in Appx.~\ref{AppxA}.

Table I shows available simulated
data~\cite{Avra} 
for the properties of Eden trees\cite{Nakan,Reis} and DLA clusters\cite{MeakinDLA,Jacobs}.  
The resistance of loopless fractals is proportional to the length of the shortest path $\ell$ between two sites which scales \cite{Avra} as $\ell \sim R^{d_{min}}$ ( so $\zeta = d_{min}$).  
Eq. (2) is seen to provide a good estimate of $\zeta$ for DLA and Eden trees in three dimensions (thus contrasting with the fractal Einstein relation).  In two dimensions, Eq. (2) is superior to the fractal Einstein relation for Eden trees, whereas for DLA both Eq. (2) and the fractal Einstein relation have a similar level of accuracy and are consistent with $\zeta = 1$.  Data for Eden trees were obtained for relatively small clusters, and it would be useful to reconsider the calculations.
\begin{table}
\caption{Equation (2) provides a significantly better estimate of $\zeta$ than the fractal Einstein relationship for Eden trees and three dimensional DLA clusters. The table is adapted from 
Ref.~\cite{Avra} using data from Refs.~\cite{Jacobs}, \cite{MeakinDLA}, \cite{Nakan}, \cite{Reis}.}
\begin{tabular}{|r|r|r|r|r|r|r|} 
\hline
Fractal & $\bar{d}$ & $d_w$ & $d_f$ & $\zeta = d_{min}$ & $\zeta = d_w - d_f$ & Eq.~(2) \\
\hline
DLA 2D & $1.20 \pm 0.05$\cite{Jacobs} & $2.64\pm 0.05$\cite{Jacobs} & $1.70 \pm 0.02$\cite{Jacobs}  & $1.00 \pm 0.02$\cite{MeakinDLA}  & $0.94\pm 0.07$ & $1.05 \pm 0.09$ \\
\hline
DLA 3D & $1.35 \pm 0.05$\cite{Jacobs}  & $3.19\pm 0.08$\cite{Jacobs}  & $2.48 \pm 0.02$\cite{Jacobs} & $1.02 \pm 0.03$\cite{MeakinDLA}  & $0.71\pm 0.10$
& $1.04 \pm 0.10$ \\
\hline
Eden tree 2D & $1.22 \pm 0.02$\cite{Nakan}  & $2.82 \pm 0.06$\cite{Reis}  & $2$ & $1.22 \pm 0.02$\cite{Nakan}  & $0.82 \pm 0.07$ & $1.10 \pm 0.05$ \\
\hline
Eden tree 3D & $1.32 \pm 0.02$\cite{Nakan} & $3.85 \pm 0.15$\cite{Reis}  & $3$ & $1.32 \pm 0.02$\cite{Nakan}  & $0.85 \pm 0.07$ & $1.31 \pm 0.09$ \\
\hline 
\end{tabular}
\end{table}

The fractal Einstein formula has been rigorously proven~\cite{Telcs2} using certain assumptions about the geometry and a technical ``smoothness" criterion on the electrostatic potential. 
In 1995, Telcs proved\cite{Telcs} that if $\zeta = d_w - d_f$  then $\bar{d} = 2d_f/d_w$ for loopless ``smooth" networks with $\zeta > 0$; however, the precise connection for general networks has not been established.  The result in Eq. (2) provides this link: it shows that $\bar{d} = 2d_f/d_w$ if, and only if, $\zeta = d_w - d_f$, irrespective of the presence of loops or the sign of $\zeta$.

A technical explanation of the assumptions underlying the Alexander-Orbach and fractal Einstein laws, and why they do not hold for the fractal tree, is given in Appx.~\ref{AppxB}). In summary,
both results implicitly assume that all sites at a distance $r$ from the origin are approximately explored uniformly by a random walker. Although the network may be spatially anisotropic, the probability fields, and hence electrostatic fields, are isotropic {\em on the network}. For the fractal tree which violates both laws, the random walker probability (and hence potential and current) fields are anisotropic {\em on the network}. This implies that a random walker shows a preference for certain directions.  Eq.~(2) does not require an isotropy condition because the electrostatic and probability fields (and hence $\zeta$, $d_w$ and $\bar d$) are similarly affected by anisotropy (Appx.~\ref{AppxD}). Figure 2 depicts the current and potential distribution on a DLA cluster. Our results indicate that the failure of the Alexander-Orbach and fractal Einstein laws are linked to the dramatic non-uniformities in the potential field and current distribution of the network.
\begin{figure}
\includegraphics[width = 0.8\textwidth]{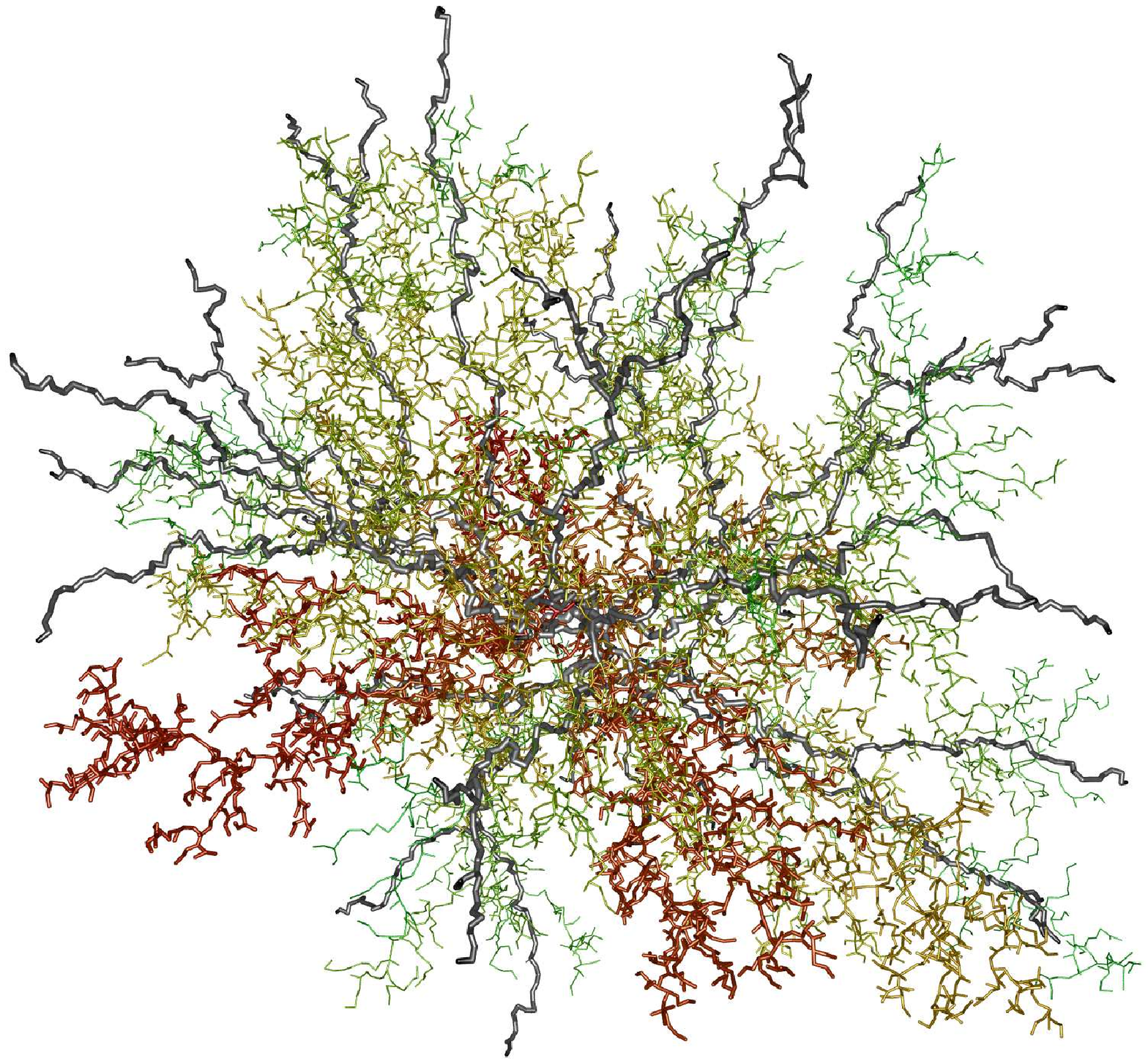}
\caption{Electrostatic fields on a diffusion limited aggregation cluster. The gray branches carry current and the potential of the non-current carrying branches is given by their colour. The red branches, which emanate near the origin, have the highest potential.  Outer branches
with potential less than 10\% of the maximum have been eliminated for clarity. The image was produced using specialised software\cite{Bourke}.}
\label{dla}
\end{figure} 

In summary, we derive a third fundamental law interrelating the properties of networks. This establishes a simple and direct connection between the Alexander-Orbach and fractal Einstein laws; if one holds, then so does the other. Because the derivation does not require the network to have a fractal dimension, we propose that the relationship holds for inhomogeneous networks \cite{Cassi1} (i.e., networks which have no $d_f$). Two examples are provided in the Appx.~\ref{AppxC}). Note that many results (e.g.~refs.\cite{ProcPRL,Condamin2}) for fractal networks are derived using the fractal Einstein or Alexander-Orbach formula.  This limits their  application  to networks, such as percolation clusters or Sierpinski graphs, on which these relationships are known to hold.  In addition to advancing our understanding of diffusion on fractals, we believe Eq.(2) will find direct application in the fields of science that rely on fractal network models.

\clearpage
\appendix
\section{Computational verification of our principal results}\label{AppxA}

Computations of $\bar{d}$ and $d_w$ are depicted in Figure 1. The results are in excellent agreement with
the analytic results. To verify Equation (2) $\rho(R)$ was computed directly for the cases $(u_1,u_2) = (1,20)$ and $(u_1,u_2) = (1,3)$.  Fits to the plots (Fig. 2) of $\rho(R)$ are in very good agreement with the analytic result $\zeta = 2- \bar{d}$.  The scaling relationship is also confirmed on a network with $\zeta > 0$   obtained by taking $(u_1,u_2) = (1,1)$ and quadrupling the branch lengths at each iteration (Haynes and Roberts, in preparation). 
In addition Eq.(2) is analytically confirmed (Haynes and Roberts, in preparation) for a non-trivial generalisation of the tree obtained by replacing the four elemental branches (of length $2^n$) of the $n$th generation of the network by a deterministic 
tree\cite{Avra} of iteration $n$ (using a T shaped generator).  In this case, $\zeta$ is unchanged because the dangling branches of the T-fractal do not contribute to the resistance, but both $d_w = \ln(6)/\ln(2)$ and $\bar{d} \neq 2d_f/d_w$ differ from the original tree.  
\begin{figure}
\includegraphics[width = 0.65\textwidth]{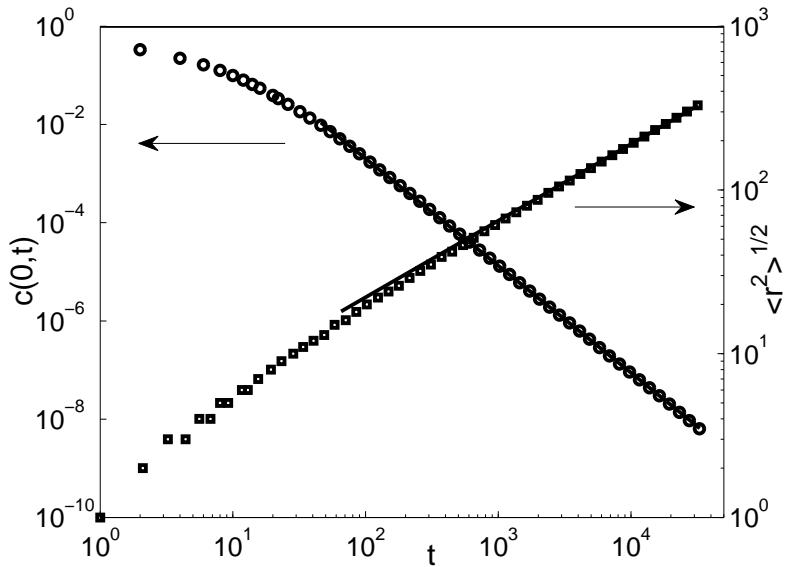}
\caption{The spectral and random walk dimensions. The probability that a walker revisits its origin at time $t$ (left hand axis); we calculate $\bar{d} = 4.37$  (Best Fit: $4.36$). On the right hand axis we plot the characteristic distance travelled by the walker $\sqrt{r^{2}(t)} \sim t^{1/d_w}$ against $t^{1/2}$  confirming $d_w = 2$.}
\label{fig:fig2}
\end{figure} 
\begin{figure}
\includegraphics[width = 0.65\textwidth]{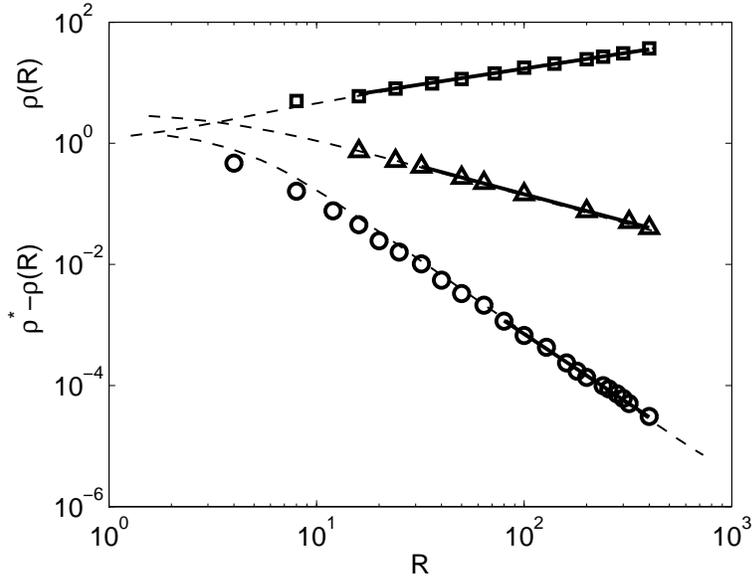}
\caption{Comparison of electrostatic and diffusive properties. The computed resistivity $\rho(R)$  of three networks (symbols) alongside the concentration $C(0,(R/b)^{d_w})$  (dashed lines) at the origin of the "constant source" diffusion problem. The value of $b$ (chosen by eye) shifts the lines horizontally. The solid lines represent lines of best fit to $\rho(R)$ : ($\circ$) The network with $(u_1,u_2) = (1,20)$, $\zeta = -2.37$,  (Best Fit:$-2.28$  ), $b = 0.9$ ; ($\vartriangle$)  The network with  $(u_1,u_2) = (1,3)$, $\zeta = -0.941$   (Best Fit:$-0.92$), $b=1.1$  ; ($\square$) A network with $\zeta >0$    discussed in the text, $\zeta = 0.5$  (Best Fit:$0.527$  ), $b = 1.1$ . 
}
\label{fig:fig3}
\end{figure} 
\section{Isotropy and anisotropy on the network}\label{AppxB}

In the main text it is demonstrated that the Alexander-Orbach (AO) and fractal Einstein (FE) laws do not apply to certain fractal networks. A new accurate formula for the resistivity exponent is derived. Both findings can be explained heuristically by considering the role of anisotropy on the network.

The probability  that a walker released at $\br=0$ at $t=0$ will be at the point $\br$ on a network after time $t$ will depend on the direction of $\br$ as well as its magnitude $r=|\br|$.  This probability is denoted as $c_a(\br,t)$, where the subscript $a$ (anisotropic) differentiates it from the function $c(r,t)$ used in the main text. The two functions can be related by
\begin{equation}
c(r,t) = \frac 1{S(r)} \int_S c_a(\br ,t) dS,
\end{equation}
where $S(r) \sim r^{df-1}$ is the area (mass) of the fractal at radius $r$.  $c(r,t)$ can be regarded as a {\em network}-spherical average, because the average on the shell is only taken over the regions occupied by the network. Equivalently, it can be called\cite{ProcPRL} the average probability {\em per site}.

\begin{figure}
\includegraphics[width = 0.65\textwidth]{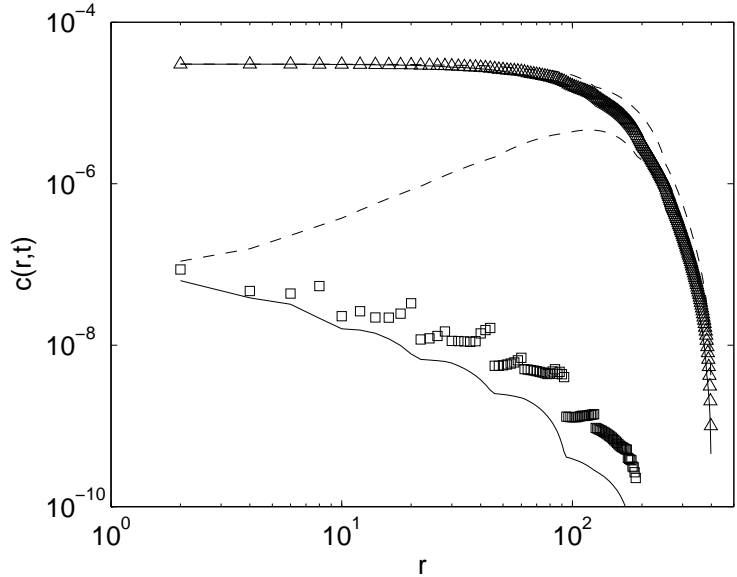}
\caption{The spherically averaged functions $c(r,t)$ for the fractal tree with $(u_1,u_2)=(1,2)$ ($\vartriangle$) and  
$(u_1,u_2)=(20,1)$ ($\square$). The lines indicate the highest and lowest values of $c_a(\br,t)$ at each $r$. The significant anistropy in the latter model is connected to the breakdown of the AO and FE laws.}
\label{anis}
\end{figure}

After time $t$, a walker released from the origin will on average have explored a region
of radius $R \sim t^{1/d_w}$. As there are $V(r)\sim R^{d_f}$ sites within
that radius, $\int_{V(R)} c_a(\br,t) dV =P$ where $P$ is the probability that the walker is in the central region. If this region is explored approximately uniformly\cite{Avra}, then
$c_a(\br,t)/c_a(0,t)$ is a slowly varying function for $r<R$.  The long time behaviour of the probability is found by setting $c_a(\br,t)\approx c_a(0,t)$ which gives $c_a(0,t) \int_{V(R)} dV \approx P $ or  $c(0,t) \approx P/V(R(t)) \sim t^{-d_f/d_w}$. This provides the rationale behind the AO law $\bar{d}=2d_f/d_w$. 
The derivation assumes that the volume $V(r(t))$ is approximately uniformly explored for $|\br|<R$. In particular, this requires that $c_a(\br,t) \approx c(r,t)$, i.e., the concentration field is approximately isotropic {\em on the network} (or network-spherically symmetric). If this is not true, the volume explored by the walker will generally not be $V(R(t))$.  Data shown in Fig.~\ref{anis} confirms that $c_a(\br,t) \approx c(r,t)$ for the fractal tree with $(u_1,u_2)=(1,2)$. This tree's  properties follow the AO law exactly. In contrast $c_a(\br,t)$ is seen to be strongly anisotropic for the fractal tree with $(u_1,u_2)=(20,1)$. The breakdown of the AO law is attributed to the non-uniform exploration of the network.  For these classes of fractal trees, the validity of the AO law is seen to depend on the probability density being isotropic on the network.

A similar requirement of uniformity is implicitly assumed in the derivation of the FE law. This is clearly seen in an examination of the total current flow $I$ through a shell of thickness $\Delta r$. By definition $I= ne\times dS \times \Delta r /(\Delta t)$ where $n$ is the charge carrier density, $e$ is the carrier charge, $dS$ is an element of area, and $\Delta t$ is the time it takes a charge to cross the shell.
Now the time scale for diffusing a distance $r$ is $t\sim r^{dw}$, so $\Delta t \sim r^{dw-1} \Delta r$. Summing over the total area of the shell gives
$$ I \sim \Delta r \int_S \frac{dS }{\Delta t} \approx \Delta r \frac{S}{\Delta t} \sim r^{df-dw},$$
and therefore $\rho=V/I \sim r^{df-dw}$ which proves the FE law. Although the argument assumes  ${\Delta t}$ is independent of direction, this is only strictly true if $c_a(\br,t)$ is uniform over the shell. 
Therefore, if diffusion exhibits preferential directions on the network
(as it does for the tree with $(u_1,u_2) = (20,1)$),  the
FE law will be invalid. 
 

The fractal tree provides a concrete example of the qualitative balance arguments expressed above.
As noted in the main text, the FE law is obeyed for the case $u_1 = 2u_2$. Rearranging the expressions for $\rho_1$ and $\rho_2$ gives $\rho_2u_2 - \rho_1u_1 = 2u_2 - u_1$; hence the condition $u_1 = 2u_2$ implies $\rho_2/\rho_1 = I_1/I_2 = u_1/u_2$, where $I_i$ ($i=1,2$) are the currents on each branch respectively.  As the ratio of the masses of the branches is $u_1/u_2$  , it is seen that conventional scaling holds because the mass and current on different branches extending from a node are balanced.  However, for the case $(u_1,u_2) = (1,20)$, a significant mass-current inbalance occurs; although there is twenty times more mass in branch $1$ than branch $2$, the current is only about $I_1/I_2 \approx 3$  times greater.  Although this discussion provides a useful picture of why the FE law fails for fractal trees studied here, note that Telcs\cite{Telcs2} has provided general conditions in terms of the potential field.


This simple numerical example also illustrates the breakdown of the AO law in terms of a mass-probability imbalance.  It is known that the ratio of currents $I_1/I_2=p_1/p_2$, where $p_i$ is the probability that a walker on branch $i$ never returns to its origin (i.e., it escapes).  Consider a walker at the first junction in the network.  There is $20$ times more mass in branch $1$ than branch $2$, but the probability of escaping along branch $1$, and never returning to the junction, is only about $3$ times greater than that of escaping along branch 
$2$.  The greater-than-expected return of walkers from branch $1$ increases $c(0,t)$, and qualitatively explains why $\bar{d}< 2d_f/d_w$.

\section{Inhomogeneous networks}\label{AppxC}

Some classes\cite{Cassi1} of important networks do not have a fractal dimension $d_f$ (called {\em inhomogeneous} networks). As the derivation of Eq.~(2) does not directly involve the mass, we hypothesise that it can be used to predict the properties of these networks. Technically, $d_w$ only exists for fractal networks, but an analogous exponent $\alpha$ is defined by $<r^2> \sim t^{\alpha}$  , whereby Eq. (2) becomes $\zeta = (2-\bar{d})/\alpha$.  Consider the fractal tree, which becomes inhomogeneous as $u_2 \to \infty$ ( $d_f \to \infty$ ). The argument used to show
$\zeta=2-\bar{d}$ is not altered in this limit, so Eq. (2) continues to hold.  A second example is
provided by the comb lattice\cite{Havlin} with an infinite spine and infinite teeth. This inhomogeneous network has $\bar{d} = 3/2$ and $\alpha = 1$ (since the teeth are one-dimensional).  The exponent $\zeta$ can be easily found if the teeth are folded against the spine (This will not affect $\bar{d}$ and $\alpha$). If the lattice is earthed along a line perpendicular to the spine (at node 0), the resistance between the $n$th node to the left of the line and the line is $\rho_n = 1/(1/n + 1/\rho_{n-1})$   with $\rho_1 = 1/2$. For large $n$ $\rho_n \approx \rho_{n-1}$, and the solution of the quadratic is $\rho_n = \sqrt{n}$ so $\zeta = 1/2$ which is consistent with $\zeta = (2-\bar{d})/\alpha$.  

\section{An explanation of Eq. (2)} \label{AppxD}

It is instructive to consider the relationship between the mean-squared distance and potential in terms of their connection with the anisotropic probability density $c_a(\br,t)$. 
There are numerous ways of defining resistance on a network. The point-to-shell resistance is defined by earthing all sites a distance $R$ from an origin, and applying a potential at that origin. The  resultant potential field is denoted
as $\phi(\br;R)$, hence the resistance is $\rho(R)=\phi(0;R)/I$ where $I$ is the current.
By the argument given in the main text, the potential at a site $\br$ is
\begin{equation}\label{phic}
\phi_a(\br;R) \approx \int_0^{(R/b)^dw} c_a(\br,t)dt.
\end{equation}
Taking the network average
\begin{equation}\label{relate}
\frac1S \int_S \phi_a(\br;R) dS \approx \frac 1S \int_0^{(R/b)^{dw}} \int_S c_a(\br,t) dS  dt
\end{equation}
and applying the scaling arguments gives Eq.~(2) of the main text
\begin{equation}\label{relateA}
\zeta=\frac{d_w}{2}(2-\bar{d}).
\end{equation}
The mean squared distance is directly related to the probability density by
\begin{equation}\label{relateB}
\langle r^2(t) \rangle = \int_V c_a(\br,t) r^2 dV = \int_0^\infty r^2 \int_S c_a(\br,t) dS   dr .  
\end{equation}

We believe that Eqn. (\ref{relateA}) can hold for structures which do not adhere to the AO and FE laws because it does not depend explicitly on the distribution of mass $M(\br)$. Consider the three exponents $\bar{d}$, $d_w$ (or $\alpha$) and $\zeta$. 
The probability $c(0,t)\sim t^{-\bar{d}/2}$ does not depend on the distribution of $M(\br)$ because the network can be distorted arbitrarily (if the inter-site distance is preserved) without changing $c(0,t)$ and hence $\bar{d}$. 
The exponents $d_w$ (or $\alpha$) and $\zeta$ may be affected by a spatial distortion of the network, but as both variables are ultimately linked to spherical averages of $c_a(\br,t)$ by Eqn.~(\ref{relate}) or~(\ref{relateB}), they will change in a consistent manner. In terms of the comb example given above; variation of the angle between the teeth and spine does not alter $c(0,t)$ and will affect $c_a(\br,t)$ and $\phi_a(\br,t)$ similarly.
For DLA clusters, and the fractal tree that violates the AO and FE laws, the fact that massive
portions of the structure are not in balance with the distribution of potential or probability does not disturb the relationship between $\bar{d}$, $d_w$ and $\zeta$.


\begin{thebibliography}{10}
\expandafter\ifx\csname url\endcsname\relax
  \def\url#1{\texttt{#1}}\fi
\expandafter\ifx\csname urlprefix\endcsname\relax\def\urlprefix{URL }\fi
\providecommand{\bibinfo}[2]{#2}
\providecommand{\eprint}[2][]{\url{#2}}

\bibitem{Havlin}
\bibinfo{author}{Havlin, S.} \& \bibinfo{author}{ben Avraham, D.}
\newblock \bibinfo{title}{Diffusion in disordered media}.
\newblock \emph{\bibinfo{journal}{Adv. Phys.}} \textbf{\bibinfo{volume}{51}},
  \bibinfo{pages}{187} (\bibinfo{year}{2002}).

\bibitem{Sahim}
\bibinfo{author}{Sahimi, M.}
\newblock \bibinfo{title}{Flow phenomena in rocks: from continuum models to
  fractals, percolation, cellular automata, and simulated annealing.}
\newblock \emph{\bibinfo{journal}{Rev. Mod. Phys.}}
  \textbf{\bibinfo{volume}{65}}, \bibinfo{pages}{1393--1534}
  (\bibinfo{year}{1993}).

\bibitem{Condamin2}
\bibinfo{author}{Condamin, S.}, \bibinfo{author}{Benichou, O.},
  \bibinfo{author}{Tejedor, V.}, \bibinfo{author}{Voituriez, R.} \&
  \bibinfo{author}{Klafter, J.}
\newblock \bibinfo{title}{First-passage times in complex scale-invariant
  media}.
\newblock \emph{\bibinfo{journal}{Nature}} \textbf{\bibinfo{volume}{450}},
  \bibinfo{pages}{77} (\bibinfo{year}{2007}).

\bibitem{Gallos}
\bibinfo{author}{Gallos, L.~K.}, \bibinfo{author}{Song, C.},
  \bibinfo{author}{Havlin, S.} \& \bibinfo{author}{Makse, H.~A.}
\newblock \bibinfo{title}{Scaling theory of transport in complex biological
  networks.}
\newblock \emph{\bibinfo{journal}{Proc. Natl. Acad. Sci}}
  \textbf{\bibinfo{volume}{104}}, \bibinfo{pages}{7746--7751}
  (\bibinfo{year}{2007}).

\bibitem{Song}
\bibinfo{author}{Song, C.~M.}, \bibinfo{author}{Havlin, S.} \&
  \bibinfo{author}{Makse, H.~A.}
\newblock \bibinfo{title}{Self-similarity of complex networks}.
\newblock \emph{\bibinfo{journal}{Nature}} \textbf{\bibinfo{volume}{433}},
  \bibinfo{pages}{392} (\bibinfo{year}{2005}).

\bibitem{OrbachRev}
\bibinfo{author}{Orbach, R.}
\newblock \bibinfo{title}{Dynamics of fractal networks}.
\newblock \emph{\bibinfo{journal}{Science}} \textbf{\bibinfo{volume}{231}},
  \bibinfo{pages}{814--819} (\bibinfo{year}{1986}).

\bibitem{Toulouse}
\bibinfo{author}{Rammal, R.} \& \bibinfo{author}{Toulouse, G.}
\newblock \bibinfo{title}{Random-walks on fractal structures and percolation
  clusters.}
\newblock \emph{\bibinfo{journal}{J. Phys. Lett. (Paris)}}
  \textbf{\bibinfo{volume}{44}}, \bibinfo{pages}{L13} (\bibinfo{year}{1983}).

\bibitem{Hughes2}
\bibinfo{author}{Hughes, B.~D.}
\newblock \emph{\bibinfo{title}{Random walks and random environments. Volume
  2}} (\bibinfo{publisher}{Clareton Press}, \bibinfo{address}{Oxford},
  \bibinfo{year}{1996}).

\bibitem{Avra}
\bibinfo{author}{ben Avraham, D.} \& \bibinfo{author}{Havlin, S.}
\newblock \emph{\bibinfo{title}{Diffusion and reactions in fractals and
  disordered systems}} (\bibinfo{publisher}{Cambridge Univ. Press},
  \bibinfo{address}{Cambridge, uK}, \bibinfo{year}{2000}).

\bibitem{MeakinDLA}
\bibinfo{author}{Meakin, P.}, \bibinfo{author}{Majid, I.},
  \bibinfo{author}{Havlin, S.}, \bibinfo{author}{Stanley, H.~E.} \&
  \bibinfo{author}{Witten, T.~E.}
\newblock \bibinfo{title}{Topological properties of diffusion limited
  aggregation and cluster-cluster aggregation}.
\newblock \emph{\bibinfo{journal}{J. Phys. A}} \textbf{\bibinfo{volume}{17}},
  \bibinfo{pages}{L975--L981} (\bibinfo{year}{1984}).

\bibitem{Orbach}
\bibinfo{author}{Alexander, S.} \& \bibinfo{author}{Orbach, R.}
\newblock \bibinfo{title}{Density of states on fractals: fractons}.
\newblock \emph{\bibinfo{journal}{J. Phys. (Paris) Lett.}}
  \textbf{\bibinfo{volume}{19}}, \bibinfo{pages}{L625} (\bibinfo{year}{1982}).

\bibitem{Mandelbrot}
\bibinfo{author}{Mandelbrot, B.~B.}
\newblock \emph{\bibinfo{title}{The fractal geometry of nature}}
  (\bibinfo{publisher}{Freeman}, \bibinfo{address}{San Francisco},
  \bibinfo{year}{1982}).

\bibitem{Jacobs}
\bibinfo{author}{Jacobs, D.~J.}, \bibinfo{author}{Mukherjee, S.} \&
  \bibinfo{author}{Nakanishi, H.}
\newblock \bibinfo{title}{Diffusion on a {DLA} cluster in two and three
  dimensions.}
\newblock \emph{\bibinfo{journal}{J. Phys. A: Math. Gen.}}
  \textbf{\bibinfo{volume}{27}}, \bibinfo{pages}{4341--4350}
  (\bibinfo{year}{1994}).

\bibitem{Witten}
\bibinfo{author}{Witten, T.~A.} \& \bibinfo{author}{Sander, L.~M.}
\newblock \bibinfo{title}{Diffusion-limited aggregation, a kinetic critical
  phenomenon.}
\newblock \emph{\bibinfo{journal}{Phys. Rev. Lett.}}
  \textbf{\bibinfo{volume}{47}}, \bibinfo{pages}{1400--1403}
  (\bibinfo{year}{1981}).

\bibitem{Doyle}
\bibinfo{author}{Doyle, P.~G.} \& \bibinfo{author}{Snell, J.~L.}
\newblock \emph{\bibinfo{title}{Random walks and electric networks.}}
  (\bibinfo{publisher}{Math. Assoc. of America}, \bibinfo{address}{Washington,
  DC, USA}, \bibinfo{year}{1984}).

\bibitem{Burioni3}
\bibinfo{author}{Burioni, R.} \& \bibinfo{author}{Cassi, D.}
\newblock \bibinfo{title}{Spectral dimension of fractal trees}.
\newblock \emph{\bibinfo{journal}{Phys. Rev. E}} \textbf{\bibinfo{volume}{51}},
  \bibinfo{pages}{2865} (\bibinfo{year}{1995}).

\bibitem{Kron}
\bibinfo{author}{Kron, B.} \& \bibinfo{author}{Teufl, E.}
\newblock \bibinfo{title}{Asymptotics of the transition probabilities of the
  simple random walk on self-similar graphs}.
\newblock \emph{\bibinfo{journal}{Trans. Amer. Math. Soc.}}
  \textbf{\bibinfo{volume}{356}}, \bibinfo{pages}{393} (\bibinfo{year}{2004}).

\bibitem{Tony}
\bibinfo{author}{Haynes, C.~P.} \& \bibinfo{author}{Roberts, A.~P.}
\newblock \bibinfo{title}{Spectral dimension of fractal trees}.
\newblock \emph{\bibinfo{journal}{Phys. Rev. E (accepted)}}
  (\bibinfo{year}{2008}).

\bibitem{Hattori}
\bibinfo{author}{Hattori, K.}, \bibinfo{author}{Hattori, T.} \&
  \bibinfo{author}{Watanabe, H.}
\newblock \bibinfo{title}{Gaussian field theories on general networks and the
  spectral dimensions}.
\newblock \emph{\bibinfo{journal}{Progr. Theoret. Phys. (Suppl.)}}
  \textbf{\bibinfo{volume}{92}}, \bibinfo{pages}{108} (\bibinfo{year}{1987}).

\bibitem{Nakan}
\bibinfo{author}{Nakanishi, H.} \& \bibinfo{author}{Herrmann., H.~J.}
\newblock \bibinfo{title}{Diffusion and spectral dimension on {E}den tree.}
\newblock \emph{\bibinfo{journal}{J. Phys. A: Math. Gen.}}
  \textbf{\bibinfo{volume}{26}}, \bibinfo{pages}{4513--4519}
  (\bibinfo{year}{1993}).

\bibitem{Reis}
\bibinfo{author}{Reis, F. D. A.~A.}
\newblock \bibinfo{title}{Scaling for random walks on {E}den trees}.
\newblock \emph{\bibinfo{journal}{Phys. Rev. E.}}
  \textbf{\bibinfo{volume}{92}}, \bibinfo{pages}{3079--3081}
  (\bibinfo{year}{1996}).

\bibitem{Telcs2}
\bibinfo{author}{Telcs, A.}
\newblock \bibinfo{title}{Random walks on graphs, electric networks and
  fractals.}
\newblock \emph{\bibinfo{journal}{Probab. Th. Rel. Fields}}
  \textbf{\bibinfo{volume}{82}}, \bibinfo{pages}{435} (\bibinfo{year}{1989}).

\bibitem{Telcs}
\bibinfo{author}{Telcs, A.}
\newblock \bibinfo{title}{Spectra of graphs and fractal dimensions {II}}.
\newblock \emph{\bibinfo{journal}{J. Theoret. Probab.}}
  \textbf{\bibinfo{volume}{8}}, \bibinfo{pages}{77} (\bibinfo{year}{1995}).

\bibitem{Cassi1}
\bibinfo{author}{Cassi, D.} \& \bibinfo{author}{Regina, S.}
\newblock \bibinfo{title}{Random walks on bundled structures}.
\newblock \emph{\bibinfo{journal}{Phys. Rev. Lett.}}
  \textbf{\bibinfo{volume}{76}}, \bibinfo{pages}{2914} (\bibinfo{year}{1996}).

\bibitem{ProcPRL}
\bibinfo{author}{O'Shaughnessy, B.} \& \bibinfo{author}{Procaccia, I.}
\newblock \bibinfo{title}{Analytical solutions for diffusion on fractal
  objects}.
\newblock \emph{\bibinfo{journal}{Phys. Rev. Lett.}}
  \textbf{\bibinfo{volume}{54}}, \bibinfo{pages}{455} (\bibinfo{year}{1985}).

\bibitem{Bourke}
\bibinfo{author}{Bourke, P.~B.}
\newblock \bibinfo{title}{Constrained diffusion limited aggregation in 3
  dimensions}.
\newblock \emph{\bibinfo{journal}{Comp. Graphics.}}
  \textbf{\bibinfo{volume}{30}}, \bibinfo{pages}{640--649}
  (\bibinfo{year}{2006}).

\end{thebibliography}

\begin{itemize}
 \item {\bf Acknowledgements} Paul Bourke, University of Western Australia, produced the DLA visualisations from the authors' data.
\end{itemize}

\end{document}